\documentstyle[preprint,aps]{revtex}
\input{psfig}

\newcommand{\beq}{\begin{equation}}
\newcommand{\eeq}{\end{equation}}

\newcommand{\ket}[1]{| #1 \rangle}
\newcommand{\bra}[1]{\langle #1 |}

\newcommand{\melement}[2]{ \langle #1 | #2 | #1 \rangle}

\begin{document}
\draft
\title{Quantum Error-Correcting Codes Need Not Completely Reveal the
Error Syndrome}

\author{P.W.~Shor, AT\&T Research, Murray Hill, NJ 07974\\
        and J.A.~Smolin, University of California at Los Angeles, 
Los Angeles, CA 90024}
\maketitle

\begin{abstract}
Quantum error-correcting codes so far proposed have not
worked in the presence of noise which introduces
more than one bit of entropy per qubit sent through a
quantum channel, nor can any code which identifies the complete 
error syndrome.  We describe a code which does not find
the complete error syndrome and can be used for reliable 
transmission of quantum information through channels
which add more than one bit of entropy per transmitted bit.
In the case of the depolarizing channel our code can be used 
in a channel of fidelity .8096. The best existing code worked
only down to .8107.
\end{abstract}
\vspace{.1 in}
\pacs{PACS: 03.65.Bz, 89.70.+c}
\vspace{.2 in}
\narrowtext

Several recent papers have dealt with the topic of good quantum 
error-correcting 
codes~\cite{SG,purification,CS,Steane,LF,ekert,PVK,EPP,SB}.
All of the efficient codes completely identify what happens to 
the state as it interacts with the environment.  In other words they 
identify the exact error syndrome.
The formal conditions which any good code must satisfy 
(see~\cite{EPP}) are less restrictive, though some have conjectured 
that error-correcting codes
must indeed identify the complete error syndrome.  There 
are trivial codes which do not gain full knowledge about
the error syndrome, for example any of the codes which do identify 
the error syndrome can be supplemented by an additional quantum system
about which no information is sought or gained.  Such examples
are trivial since the additional system is in a product state 
with the system which is actually involved in the coding and it is 
clear that such a code can only be less efficient than the codes 
from which they are derived.  In \cite{EPP}, a ``hashing''
code is presented which, while it does not completely identify 
the error syndrome, achieves precisely the same rate as the 
``breeding'' protocol of
\cite{purification,EPP} which does.  Here we present a non-trivial 
code 
which does not identify the entire error syndrome {\em and} can 
work in a 
noisier channel than any code which does.  

The typical error model used in analyzing quantum error-correcting codes is
that of independent depolarization.  In terms
of the probability $x$ of not being depolarized, 
each qubit (two state quantum system) which is sent through a 
channel has a probability
$f=\frac{3x+1}{4}$ of being
transmitted untouched, and equal probabilities $(1-f)/3$ of 
1) flipping the amplitude ($\ket{\!\!\uparrow}$ vs. 
$\ket{\!\!\downarrow}$),
2) changing the sign of the relative phase of $\ket{\!\!\uparrow}$ and 
$\ket{\!\!\downarrow}$ or 3) both.
The specification of which type of error (or none) happened to each 
qubit is what is known as the
error syndrome.  Clearly, if one knew the error syndrome, all the
qubits could be corrected by simply flipping each bit's direction or phase
(or both) as needed, using the Pauli matrices.

We present our code first in the language of quantum entanglement 
purification protocols, and 
then describe the corresponding direct quantum error-correcting code.
In quantum purification protocols~\cite{purification,EPP} two-particle states 
$\ket{\Phi^+}=1/\sqrt{2} 
(\ket{\!\uparrow\uparrow}+\ket{\!\downarrow\downarrow})$
are prepared by one participant (Alice) and one of the particles is sent 
through
the channel to the other participant (Bob).  Using the four Bell states 
\begin{eqnarray}
&\ket{\Phi^\pm}=\frac{1}{\sqrt{2}}
(\ket{\!\uparrow\uparrow}\pm\ket{\!\downarrow\downarrow})&
\nonumber\\
&{\rm and}& \\
&\ket{\Psi^\pm}=\frac{1}{\sqrt{2}}
(\ket{\!\uparrow\downarrow}\pm\ket{\!\downarrow\uparrow})
& \nonumber
\end{eqnarray}
as a basis, the error model can be expressed as taking the $\ket{\Phi^+}$
states
into density matrices of the Werner form
\beq
W=\left( \begin{array}{cccc}
f& & & \\
 &\frac{1-f}{3}& & \\ 
 & &\frac{1-f}{3}& \\ 
 & & &\frac{1-f}{3}\\
\end{array}\right) \ .
\label{werner}
\eeq
$f=\melement{\Phi^+}{W}$ is then the fidelity of $W$ relative 
to $\ket{\Phi^+}$.  In this language, amplitude errors interchange
$\ket{\Phi}$ and $\ket{\Psi}$ states, and phase errors interchange
plus and minus states.

Our improved purification protocol uses the Bilateral exclusive or 
(BXOR) operation
of~\cite{purification}.  Alice and Bob each apply the exclusive or (XOR) 
operation:
\beq \begin{array}{ccr} 
U_{XOR}& = & \ket{\!\uparrow_S\uparrow_T}\bra{\uparrow_S\downarrow_T}+
         \ket{\!\uparrow_S\downarrow_T}\bra{\uparrow_S\uparrow_T}  \\
    &  & +\ket{\!\downarrow_S\downarrow_T}\bra{\downarrow_S\downarrow_T} +
         \ket{\!\downarrow_S\uparrow_T}\bra{\downarrow_S\uparrow_T}
\end{array}
\eeq
to the corresponding particles of two Bell states which have been
shared through the channel.   It can be easily seen that when 
$U_{XOR}$ is applied to two qubits, each in one of the basis
states $\ket{\!\uparrow}$ or $\ket{\!\downarrow}$, that one of them
is left alone and the other is left in the state corresponding
to the classical XOR of the two original states.  These
are called the source and target qubits respectively.  

\begin{figure}
\centerline{\psfig{figure=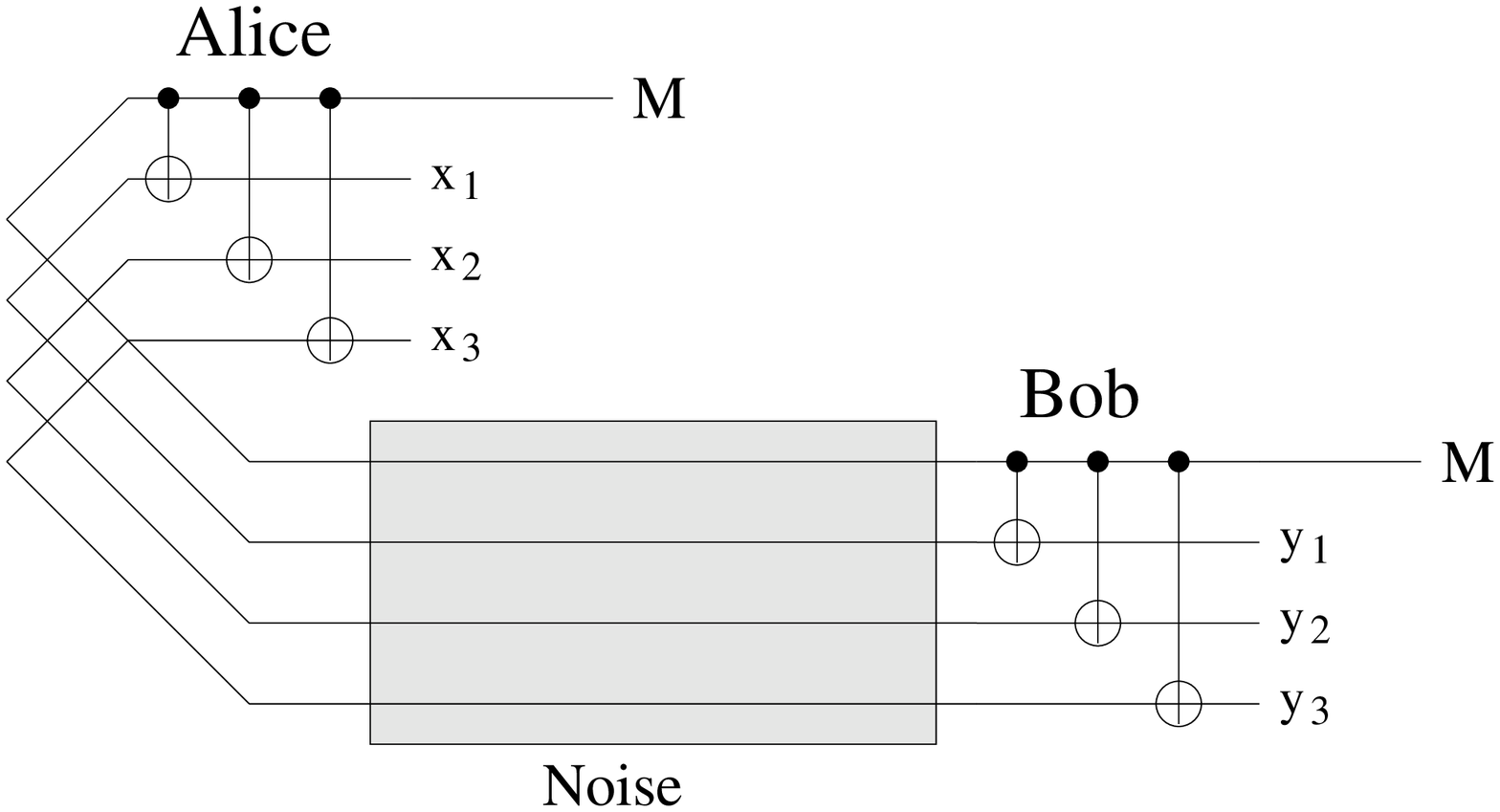,width=5in}}
\caption{The first stage of the $k=4$ code, showing the overall
purification view.
\label{ss2}}
\end{figure}

The first stage of the purification protocol is for Alice
and Bob to group their noisy pairs of particles into blocks
of size $k$.  Next they apply the BXOR operation with one pair
as the source and each of the other pairs in the block as
the target in turn.  The target qubits are all measured
in the $z$ basis and Alice sends her classical results as 
bitstring $x$ to Bob, whose results are bitstring $y$, as shown in 
Figure~\ref{ss2}.
Bob compares his results to Alice's and checks whether each bit
agrees or disagrees, which is just taking the bitwise XOR, $x \oplus y$.  
The remaining unmeasured source pair 
is then in one of $2^{k-1}$ post-selected density matrices corresponding
to the $2^{k-1}$ results of $x\oplus y$.  All are
diagonal in the Bell basis.

\begin{figure}
\centerline{\psfig{figure=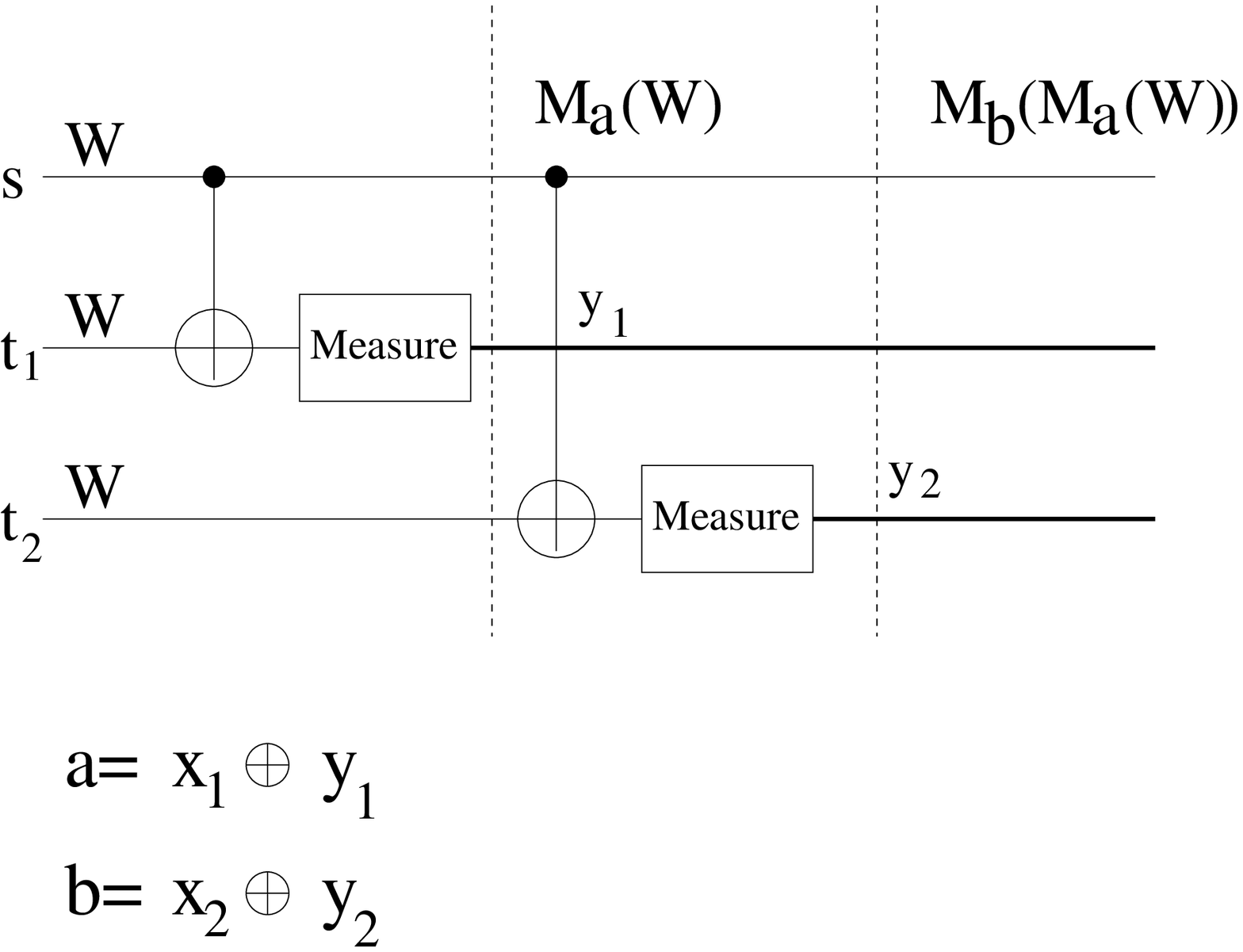,width=5in}}
\caption{Bob's view of the conditional $M$'s as the
BXORs are done in sequence.
\label{ss1}}
\end{figure}

The expected entropy of this ensemble is expressed most
simply by a recursively defined function:
\begin{eqnarray}
&& S(n,M) := \nonumber \\
&&\ {\rm if}(n==1) {\rm\ then\ return\ }( h(M) ) \nonumber \\
&&\ {\rm else\ return}\nonumber\\
&&\ \ \ p_0(M) S(n-1,M_0(M)) + p_1(M) S(n-1,M_1(M))\nonumber\\
\end{eqnarray}
where $h(M)=-{\rm Tr}(M\log M)$, $p_0(M)$ and $p_1(M)$ are the probabilities
that Alice and Bob's results with matrix M as a source and target state
$W$ will agree or disagree, and
$M_0(M)$ and $M_1(M)$ are the post-selected density matrices for the
source matrix $M$ when Alice and Bob's results agree and disagree.
Bob's view of this is shown in Figure~\ref{ss1}.
It is straightforward to calculate these functions using the facts that
the BXOR operation maps Bell states into Bell states as shown in 
Table~\ref{belltable}, that the matrices $M$ and $W$ are Bell diagonal, and
that Alice and Bob's measurements will agree when then have
$\ket{\Phi^\pm}$ and disagree when they have $\ket{\Psi^\pm}$.
We have then have for the $p$ functions
\widetext
\begin{eqnarray}
&p_0(M)=(f+g)\langle\Phi^+|M|\Phi^+\rangle + 
    2g \langle\Psi^+|M|\Psi^+\rangle +
    (f+g)\langle\Phi^-|M|\Phi^-\rangle +
    2g \langle\Psi^-|M|\Psi^-\rangle &  \nonumber\\
&p_1(M)=2g \langle\Phi^+|M|\Phi^+\rangle +
    (f+g)\langle\Psi^+|M|\Psi^+\rangle +
    2g \langle\Phi^-|M|\Phi^-\rangle +
    (f+g)\langle\Psi^-|M|\Psi^-\rangle &\end{eqnarray}
\narrowtext
\noindent and for the $M$ functions
\begin{eqnarray}
&&\langle\Phi^+|M_0(M)|\Phi^+\rangle
=\frac{f\melement{\Phi^+}{M} + g\melement{\Phi^-}{M}}{p_0(M)} \nonumber\\
&&\melement{\Psi^+}{M_0(M)}
=\frac{g\melement{\Psi^+}{M} + g\melement{\Phi^-}{M}}{p_0(M)} \nonumber\\
&&\melement{\Phi^-}{M_0(M)}
=\frac{g\melement{\Phi^+}{M} + f\melement{\Phi^-}{M}}{p_0(M)} \nonumber\\
&&\melement{\Psi^-}{M_0(M)}
=\frac{g\melement{\Psi^+}{M} + f\melement{\Psi^-}{M}}{p_0(M)} \end{eqnarray}
and
\begin{eqnarray}
&&\langle\Phi^+|M_1(M)|\Phi^+\rangle
=\frac{g\melement{\Phi^+}{M} + g\melement{\Phi^-}{M}}{p_1(M)} \nonumber\\
&&\langle\Psi^+|M_1(M)|\Phi^+\rangle
=\frac{f\melement{\Psi^+}{M} + g\melement{\Psi^-}{M}}{p_1(M)} \nonumber\\
&&\langle\Phi^-|M_1(M)|\Phi^+\rangle
=\frac{g\melement{\Phi^+}{M} + g\melement{\Phi^-}{M}}{p_1(M)} \nonumber\\
&&\langle\Psi^-|M_1(M)|\Phi^+\rangle
=\frac{g\melement{\Psi^+}{M} + f\melement{\Psi^-}{M}}{p_1(M)} 
\end{eqnarray}
where we have written $g=(1-f)/3$ for convenience.  Note that
$M_0(M)$ and $M_1(M)$ are diagonal so these equations specify
them completely.

If Alice and Bob have a large number of such results, and
when $S(k,W) < 1$, they can use the breeding purification 
method of~\cite{purification} to completely determine the error 
syndrome {\em of these remaining states}~\cite{FN}.
The complete error syndrome of the 
amplitude errors is found, but since the BXORs done within the
blocks of $k$ determine nothing about the phase errors, only
the overall phase error of the block of $k$ is determined.
This procedure will result in a yield of pure 
$\ket{\Phi^+}$ states of $\frac{1-S(k,W)}{k}$.  These can then be used
for quantum teleportation~\cite{teleportation} to transmit
qubits safely through the noisy channel.

The breeding protocol assumes Alice and Bob share a set of
unknown Bell states and a supply of $\ket{\Phi^+}$ states 
known to be pure.  If a sequence of $n/2$ Bell states
is represented by a length $n$ bitstring $x$, the parity $x \cdot s$ 
of any subset $s$ of the
bits of the string can be collected into the amplitude bit ($\ket{\Phi}$ vs.
$\ket{\Psi}$)
of one of the initially pure pairs, without disturbing the $n/2$ 
unknown Bell states.
This is accomplished by repeatedly using the BXOR operation 
with the pure pair as the target.  Each of the unknown 
states whose amplitude bit is part of $s$ is used
as a source, and each one whose phase bit is selected by 
$s$ is pre- and post-processed by the bilateral rotation of
$\pi/2$ around the $y$ axis (which has the effect of swapping
the amplitude and phase bits, and then swapping them back).
The subset parity $s$ is then determined by measuring
the target state in the $z$ basis.  

The probability of  any two strings $x$ and $x'$ having $m$ such random
subset parities all agree is $1/2^m$.  Given a random independent
noise process the original ensemble of
possible bitstrings has most of its weight in a set of ``typical'' strings 
containing $2^{\frac{n}{2}S+\delta}$ ($S$ is the entropy
per Bell state, $\delta$ is small compared to $nS$).  For such a 
distribution the 
collision probability of {\em any} string in the 
typical set other than $x$ having the same $m$ random subset 
parities is 
\beq
p_c=\frac{2^{\frac{n}{2}S+\delta}}{2^m}\ .
\eeq
The probability of $x$ falling outside of the typical set is of
order $O({\rm exp}(-\delta^2 n))$~\cite{schumacher}.  Therefore,
if $m$ is chosen slightly larger than $\frac{n}{2}S$, the
original string $x$ can be determined from the $m$ subset
parities with high probability.  All the Bell states can then
be corrected to pure $\ket{\Phi^+}$ states.  $m\approx \frac{n}{2}S$
pure $\ket{\Phi^+}$ states had to be measured in the process of
finding the $m$ subset parities, and so much be replaced, for a net
yield of $D=1-S$.

The breeding method was only shown in~\cite{purification}
to work on a single Werner channel rather than the ensemble 
resulting from our $k$-way encoding.  If Alice and
Bob simply had $k-1$ channels of different fidelities
they could clearly just use the breeding method, or any
other, on each channel separately.  However, Alice
does not know into which type of channel each pair falls.
Fortunately, the breeding protocol depends only on an
ensemble of $n$ bits having most of its weight in a 
set of ``typical'' strings containing $2^{\frac{n}{2}S+\delta}$
members, which the receiver Bob can enumerate.  
It is apparent that the individual $k-1$ channels
each have such a typical set and 
so, therefore, will the collection of all of them,
even though only Bob can determine this set.  Another 
important feature of the breeding and hashing protocols
is that Alice and Bob choose {\em randomly}
among a set of operations determined only by the channel 
fidelity.  This implies that Alice can
do her part of the procedure with no knowledge of
any sort from Bob.

Because of the formal equivalence
of measurement of half of a Bell state and preparation
of a qubit, any purification protocol requiring only 
one-way communication can be converted
into a more explicit quantum error-correcting code~\cite{EPP}.
Our protocol must work regardless of Alice's
classical measurement results within the blocks of $k$.
(Different results cannot convey any information to
Alice because her half of each pair has not even interacted
with the noise).  In particular, our protocol must work  
when Alice's results are 
all $\ket{\!\!\downarrow}$.
This result means that Bob's bits, before having been acted
on by the noise, must have been prepared all in the same state,
without specifying which state that is.  In other words,
Alice prepares a state of the form
$\frac{1}{\sqrt{2}}(\ket{\!\uparrow\uparrow\uparrow\ldots\uparrow\uparrow}+
\ket{\!\downarrow\downarrow\downarrow\ldots\downarrow\downarrow})$ and sends
$k-1$ of the bits through the channel.  Bob's half of the
BXOR operation is done as the decoding state, and amounts
to the incomplete measurement of which of the qubits have
different amplitude from the first, without determining
the actual amplitude of any of them other than relative to the first.

\begin{figure}
\psfig{figure=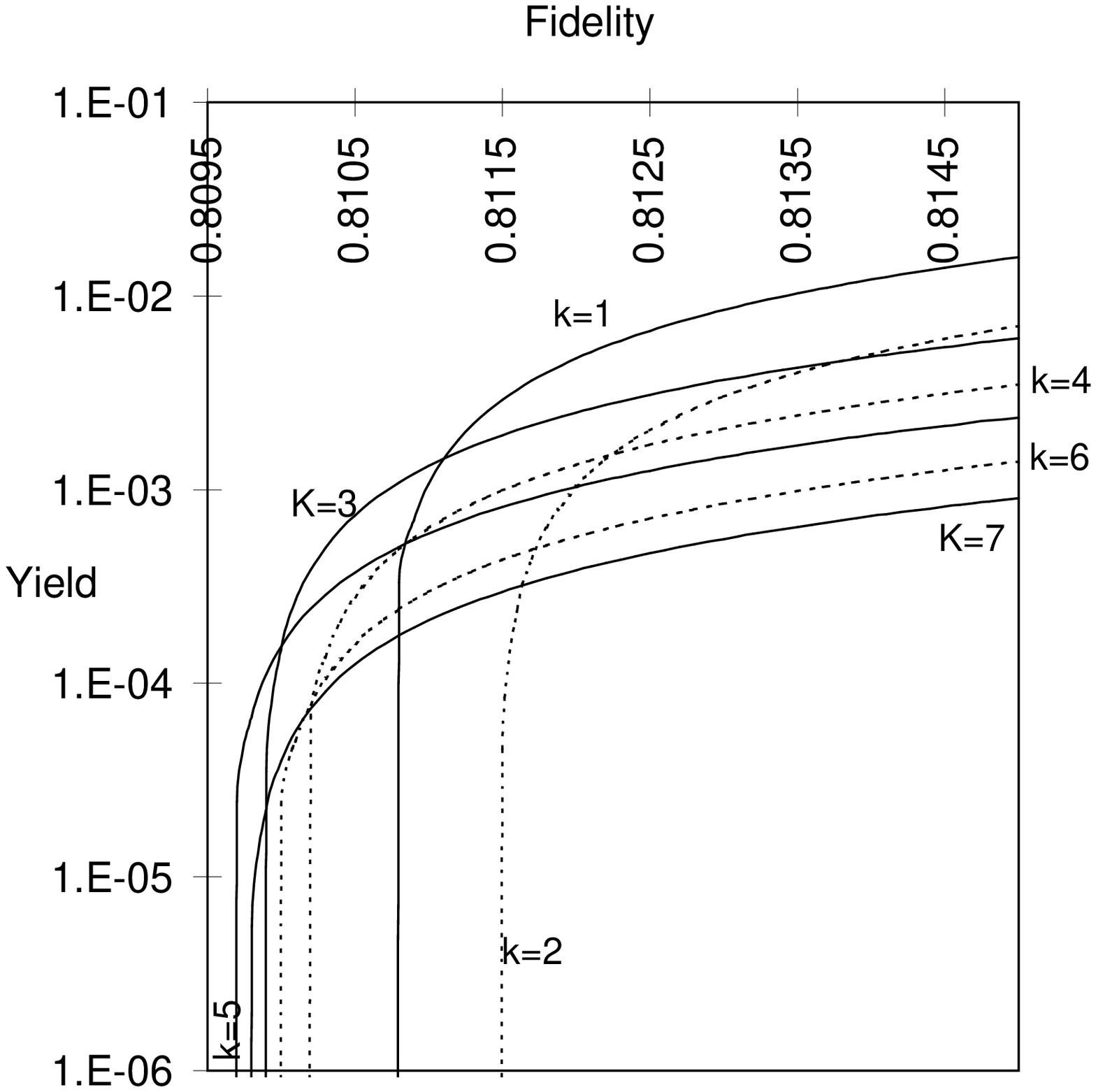,width=5in}
\caption{The yield of distillable $\ket{\Phi^+}$ states by 
purification or the fraction of transmitted bits which can be 
protected from
noise as a function of channel fidelity for various
values of $k$.  Note that the curves are all in order from
$k=1$ to $k=7$ along the right side of the graph.
\label{plot}}
\end{figure}

The hashing method applied directly to the states $W$ (the $k=1$ case)
determines the full error syndrome, and allows error correction in
channels of fidelity where $h(W_f)<1$.  $h(W_f)=1$ for $f=.8107$.  Our
new method extends this to as low as $f=.8096$ for $k=5$.  Other
values are given in the Table~\ref{ktab}.  The fraction $D$ of the
bits transmitted through a channel which can be protected for a given
channel fidelity is plotted in Figure~\ref{plot} for $k=1$~to~$7$.  It
is not yet known what the minimum fidelity channel is which can still
have some capacity for transmission of undisturbed qubits, and our
result only improves the previously known result by about 0.1\%.  It
is known, and proved in~\cite{EPP}, that channels of fidelity $f\leq
5/8$ have no capacity.  There is still obviously a lot of room between
that result and ours.  Our result demonstrates that quantum
error-correcting codes do not need to find the whole error syndrome, a
property that any lower bound on the fidelity of a channel which can
transmit undisturbed qubits must share.

It should be noted as well that this and other protocols which
are designed to work on depolarizing Werner channels 
will work on any noise which acts independently on each particle
transmitted through the channel and which turns $\ket{\Phi^+}$'s into
density matrices satisfying
\beq
g={\rm Max}(\melement{\zeta}{\rho})\geq f_c
\label{spin}
\eeq
where the maximum is found over all maximally entangled
four-dimensional $\ket{\zeta}$ and  
$f_c$ is the cutoff fidelity above which the code would 
work in a depolarizing channel.  This is seen by rotating 
$\zeta_{max}$
into the direction of $\ket{\Phi^+}$ which can be done by entirely local
actions (\cite{Gisin}) and then randomly rotating the state
by applying a randomly selected $SU(2)$ to both Alice's and Bob's
particles (this is the random bilateral rotation procedure 
of~\cite{purification}, explained in more detail in~\cite{EPP}).  This
results in a Werner density matrix of fidelity $f=g$ given by Eq.~\ref{spin}.

\begin{table}[p]
\begin{tabular}{r|llll|l}
 &\multicolumn{4}{c|}{source}& \\
target&$\Psi^-$&$\Phi^-$&$\Phi^+$&$\Psi^+$\\
\cline{1-6}
 &$\Psi^+$&$\Phi^+$&$\Phi^-$&$\Psi^-$&(source)\\
$\Psi^-$&$\Phi^-$&$\Psi^-$&$\Psi^-$&$\Phi^-$&(target)\\
 \cline{1-6}
&$\Psi^+$&$\Phi^+$&$\Phi^-$&$\Psi^-$&(source)\\
$\Phi^-$&$\Psi^-$&$\Phi^-$&$\Phi^-$&$\Psi^-$&(target)\\
 \cline{1-6}
&$\Psi^-$&$\Phi^-$&$\Phi^+$&$\Psi^+$&(source)\\
$\Phi^+$&$\Psi^+$&$\Phi^+$&$\Phi^+$&$\Psi^+$&(target)\\
 \cline{1-6}
&$\Psi^-$&$\Phi^-$&$\Phi^+$&$\Psi^+$&(source)\\
$\Psi^+$&$\Phi^+$&$\Psi^+$&$\Psi^+$&$\Phi^+$&(target)\\
\end{tabular}
\caption{The BXOR mappings of Bell states onto Bell states
\label{belltable}}
\end{table}

\begin{table}
\begin{tabular}{lll|ll}
k&f&&k&f\\
\cline{1-5}
1&.8107&&8&.8101\\
2&.8115&&9&.8101\\
3&.8099&&10&.8103\\
4&.8101&&11&.8104\\
5&.8096\ \ \ Best&&12&.8106\\
6&.8100&&13&.8107\\
7&.8098&&14&.8108\\
\end{tabular}
\caption{The value of fidelity $f$ for which
$S(k,f)=1$.   Values of $k$ not shown all
work less well than the direct hashing method ($k=1$).
\label{ktab}}
\end{table}

\widetext
\vspace{1 in}

\begin{thebibliography}{99}
\bibitem{FN} The more explicitly one-way hashing protocol 
of~\protect\cite{EPP} could also be used.  It does not completely 
determine the error syndrome of the states on which is is employed, 
but is effectively the same as the breeding method as all pieces of 
the error-syndrome which are not determined are essentially 
discarded for a net zero gain.
\end{thebibliography}
\end{document}